\documentclass[prl,aps,superscriptaddress,twocolumn]{revtex4-1}
\usepackage{graphicx} 
\usepackage{subfigure} %
\usepackage{dcolumn}
\usepackage{bm}
\usepackage{tabularx}
\usepackage[latin1]{inputenc}
\usepackage[usenames,dvipsnames]{color}
\usepackage[mediumspace,mediumqspace,Grey,squaren]{SIunits} 
\usepackage[version=3]{mhchem} 
\usepackage{amsmath}
\usepackage{amsfonts}
\usepackage{mathrsfs}
\usepackage{amssymb} 
\def\be{\begin{equation}}
\def\ee{\end{equation}}
\def\ba{\begin{array}}
\def\ea{\end{array}}
\def\bea{\begin{eqnarray}}
\def\eea{\end{eqnarray}}

\begin{document}

\title{Topological symmetry breaking in viscous coarsening}

\date{\today}

\author{David Bouttes}
\affiliation{Laboratoire PMMH, UMR 7636 CNRS/ESPCI/Univ. Paris 6 UPMC/Univ. Paris 7 Diderot, 10 rue Vauquelin, 75231 Paris cedex 05, France}

\author{Emmanuelle Gouillart}
\affiliation{Surface du Verre et Interfaces, UMR 125 CNRS/Saint-Gobain, 93303 Aubervilliers, France}

\author{Damien Vandembroucq}
\affiliation{Laboratoire PMMH, UMR 7636 CNRS/ESPCI/Univ. Paris 6 UPMC/Univ. Paris 7 Diderot, 10 rue Vauquelin, 75231 Paris cedex 05, France}


\begin{abstract}
The crucial role of hydrodynamic pinch-off instabilities is evidenced
in the coarsening stage of viscous liquids. The phase
separation of a barium borosilicate glass melt is studied by in-situ
synchrotron X-Ray tomography at high temperature. The high viscosity contrast between the less viscous phase and the more viscous phase induces a
topological symmetry breaking: capillary breakups occur preferentially
in the less viscous phase. As a result, contrasting morphologies are obtained in the two phases. This symmetry breaking is illustrated on three
different glass compositions, corresponding to different volume
fractions of the two phases. In particular, a fragmentation phenomenon,
reminiscent of the end-pinching mechanism proposed by Stone et
al.~\cite{Stone-JFM86,Stone-JFM89} is evidenced in the less viscous
phase.
\end{abstract}

\maketitle

Understanding the fragmentation of liquids is a long-standing problem
in fluid dynamics~\cite{Villermaux-ARFM07}. The competition between
surface tension, inertial and viscous effects at play in droplet
generation and spray formation is of paramount importance in fields
such as microfluidics~\cite{Herminghaus-RPP12}, ocean-atmosphere
exchanges~\cite{Veron-ARFM15} or volcanic
eruptions~\cite{Villermaux-CRM12,Gonnermann-AREPS15}.
Of particular interest for industrial applications
is the size distribution arising
from the fragmentation process~\cite{Villermaux-ARFM07}.

In the context of viscous coarsening of phase separated silicate melts,
we have recently evidenced an original fragmentation phenomenon
leading to a power-law size distribution of
droplets~\cite{BGBDV-PRL14,Bouttes-ActaMater15}. Here the scale-free
character of the fragment-size distribution is a direct
inheritance of the self-similar structure of the inter-connected
cluster.
In the course of liquid-liquid phase separation, interface tension driven
coarsening induces a temporal growth of the characteristic size of the
phase domains $\ell(t)\propto t^{\alpha}$ where the exponent $\alpha$
depends on the transport mechanism at play (diffusion,
advection...)~\cite{Bray2003}. A major feature of the coarsening
phenomenology is thus the observation of self-similar structures obeying
dynamical scaling invariance: structures are left statistically
invariant after renormalization by the characteristic length scale
$\ell(t)$.

In these experiments, as in other works, dynamical scaling is a
powerful tool to describe the statistical features of the coarsening
\cite{Bray-AdvPhys94,Cugliandolo-PhysicaA10}, and can be derived in a
few situations \cite{Sicilia-PRE07}.  The starting point of the
scaling laws governing domain growth is often the understanding of the
local mechanisms at play, e.g. hydrodynamic pinch-off for viscous
coarsening.  The importance of these local mechanisms was stressed by
Siggia in his pioneering work~\cite{Siggia-PRA79}, but they have
received little
attention~\cite{Aarts-NJP05,Voorhees-NatPhys10,Royall-PRL15}, despite
being key to the specific geometrical features of the coarsening, such
as the fragmentation we observed.

Here we show that a viscosity contrast 
can break the symmetry between the two phases and lead to a different
morphology of the two phases, and to the
fragmentation of the less viscous phase.  We use \emph{in situ} synchrotron
microtomography to follow the temporal development of viscous
coarsening in phase-separated barium borosilicate melts at high
temperature, with sufficient spatial and temporal
resolution to access the fine details
of the evolution of the structure of the two liquids. In particular, we
unveil the crucial importance of the hydrodynamic pinch-off mechanisms
at play in viscous coarsening and the effect of a viscosity contrast.

{\it Phase-separated barium-borosilicate glasses} \--- In the
following we present results obtained on three glasses
hereafter denoted by G$_1$, G$_2$, G$_3$ (see compositions in Table
\ref{compos}). These glasses lie on the same tie-line: when heated in the
range $T=1160^\circ {\mathrm C} - 1210^\circ {\mathrm C}$ used in the
present series of experiments, they decompose into a {\it viscous}
silica-rich phase and a {\it fluid} barium-rich phase. The viscosity
contrast between the two phases is about 5 orders of
magnitudes~\cite{Bouttes-ActaMater15}. Within the limits of the
precision of the formulation and the control and variarions of
temperatures, the composition of the two separated phases obtained in
the different experiments are extremely
similar~\cite{Bouttes-Thesis14}. As summarized in
Table~I, only the volume fraction of the two phases
depends on the initial composition. Here the volume fractions of the
fluid phase are respectively $\Phi_1= 0.27$,
$\Phi_2 = 0.45$ and $\Phi_3=0.72$. For the three volume fractions, the
initial microstructure is bicontinuous, suggesting spinodal decomposition.

{\it In-situ experiments} \--- X-ray tomographic experiments have been
performed on beamline ID19 at the European Synchrotron Radiation Facility
(ESRF). Glass samples 2~mm in diameter were studied \emph{in situ} at high
temperatures using a dedicated furnace.  A high-flux pink beam of energy
32 keV was used, so that it took 15 s to acquire a full 3-D image,
corresponding to a volume of size $700\mu$m$\times
700\mu$m$\times350\mu$m, with voxels of size $1.1\,\mu$m. Additional
details on the experimental set-up and data processing (reconstruction,
segmentation and characterization of the 3D geometry) can be found in
Ref.~\cite{Bouttes-Thesis14,BGBDV-PRL14,Bouttes-ActaMater15}.

\begin{table}
\begin{tabular}{c|c|c|c|c|c}
Glass  &SiO$_2$ &B$_2$O$_3$ &BaO &T ($^\circ$C) &Fluid phase\\
compositions & & & & &vol. \%\\
\hline
G$_1$ &65 &17 &18 &1160 &27$\pm2$\\
G$_2$ &60 &19 &21 &1210 &45$\pm2$\\
G$_3$ &50 &22 &28 &1160 &72$\pm2$\\
Viscous phase &80 &17 &3 & &0\\
Fluid phase &36-38 &26 &36-38 & &100
\label{table-glasses}
\end{tabular}

\caption{Compositions in weight percentages of the three glass
  compositions G$_1$, G$_2$, G$_3$ under study and
  of the two phases, (viscous) silica-rich and (fluid)
  barium-rich respectively, obtained after separation. The
  temperatures of heat treatments are also indicated.
\label{compos}}
\end{table}

{\it Coarsening results from series of topological events} \--- Coarsening
studies emphasize the scaling properties of domain growth
and  usually ignore the details of the geometrical changes  at
play at the local scale.  Independently of the nature of the transport mechanism
(diffusion or advection), statistical invariance
upon time has important consequences from a topological point of
view. The characteristic length scale of the structure can be
interpreted as a typical mesh size and the growth of this length scale
implies a decrease of the number of loops of the structure, hence a
decrease of the topological genus of the structure. However it is not possible
to change the genus of a structure by a continuous deformation.
Coarsening results necessarily from a sequence of
topological events: ruptures of links and their complementaries,
resorptions of loops. This point has been recognized early on by Siggia
\cite{Siggia-PRA79}, who proposed that viscous coarsening results from
capillary breakups and retractions of ligaments.

\begin{figure}
\includegraphics[width=0.99\columnwidth]{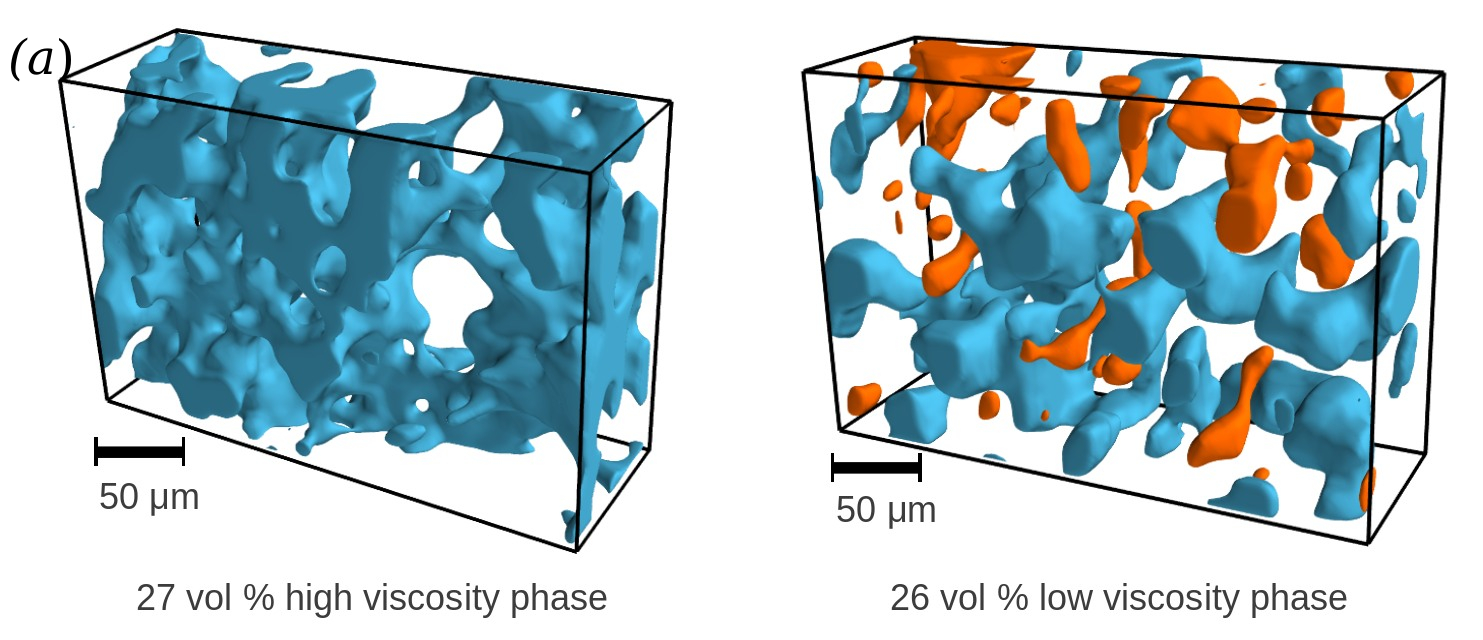}
\caption{Evolution of the topology of the minority (barium-rich) phase
  of glass G$_2$ under coarsening. Top: snapshots of an elementary
  topological event, the rupture of a capillary bridge.  Bottom:
  illustration of the same event in the complementary (silica-rich)
  phase, now the resorption of a hole.}
\label{pinch-off}
\end{figure}

\begin{figure}
\includegraphics[width=0.85\columnwidth]{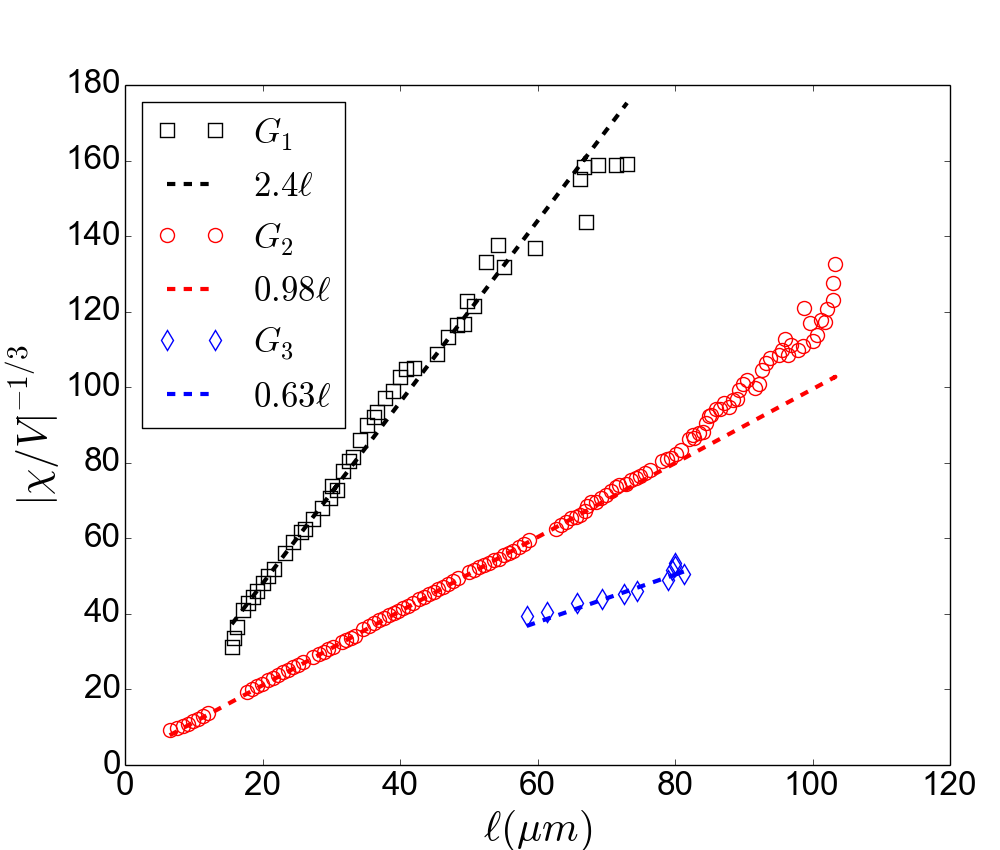}
\caption{Evolution under coarsening of the inverse cubic root of the
  volumic Euler characteristic $\chi$ of the percolating domain vs.
  characteristic size $\ell$ for the three glasses G$_1$, G$_2$ and
  G$_3$. The dotted lines show indicative linear behaviors associated
  with the expected scaling for viscous coarsening.}
\label{Euler}
\end{figure}

As illustrated on Fig.~\ref{pinch-off}(top) obtained on glass G$_2$,
such pinch-off events can be identified along the coarsening stage.  A
capillary bridge breaks up and the retraction of the resulting
ligament feeds the remaining structure\---hence the growth of its
characteristic size. As shown in Fig. \ref{pinch-off}(bottom), the
same elementary event can be seen as the retraction of a loop in the
complementary phase.

Beyond this example of topological event at local scale, the Gauss-Bonnet
  theorem gives a global mathematical characterization of the deep link
  between domain growth and
  topology~\cite{Levitz-CCR07,Levitz-EPJAP12}:
\begin{equation}
    \int_S K \mathrm{d}S = 2\pi \chi\;.
\label{Gauss-Bonnet}
\end{equation}
The integral of the Gaussian curvature $K$ on the surface of the
domains gives immediately access to the Euler characteristic $\chi$, a
topological invariant which is the sum of isolated objects
$\mathcal{N}$, minus the number of loops $\mathcal{L}$, plus the
number of cavities $\mathcal{O}$: $\chi = \mathcal{N} - \mathcal{L} +
\mathcal{O}$. During coarsening, we expect capillary bridges to break,
hence decreasing the number of loops and the absolute value of the
Euler characteristic.
The dynamic scaling invariance at play for the evolution of the
morphology of the domains thus directly translates to their
topology~\cite{Fialkowski-PRL01,Voorhees-EPL09}. Let us call $\ell(t)$
the characteristic length scale at time $t$. Dynamic invariance
imposes that the statistical distribution of Gaussian curvatures
$P[K(t)]$ obeys:
\begin{equation}
P[K(t)] = \ell(t)^2 \Psi [K(t)\ell(t)^2]\;.
\label{Gauss-curvature}
\end{equation}
Combining (\ref{Gauss-Bonnet}) and (\ref{Gauss-curvature}) results in the
following evolution of the volumetric Euler characteristic $\chi_V$:
\begin{equation}
\chi_V(t)=\frac{\chi(t)}{V} 
 \simeq \frac{S}{V} \overline{K}(t) \simeq \ell(t)^{-3}
 \simeq \left( \frac{\gamma}{\eta}t\right)^{-3}
\end{equation}
where $S$ and $V$ are the total surface and volume of the domain.
The interface tension $\gamma$ and the viscosity
$\eta$ are
associated with Siggia's scaling of viscous coarsening $\ell(t) \simeq
(\gamma/\eta)t$~\cite{Siggia-PRA79, Bouttes-ActaMater15}.

Using the algorithm of
Ref.~\cite{Nagel-JMicro-00,Lang-JMicro-01,Vogel-CG10} we computed the
evolution under coarsening of $\chi_V$ versus the expected
characteristic length scale $\ell(t)=V/S$, here computed on the
largest (interconnected) domain of the fluid phase for the
three liquids. In Fig.~\ref{Euler} we represent the
evolution of the effective length scale $\ell_\chi (t) =
\chi_V^{-1/3}(t)$ versus the coarsening characteristic length
$\ell(t)$.
The remarkable linear correlation that we obtain between the two
quantities illustrates that dynamic scaling is valid for topological
quantities (such as $\chi_V$) as well, and emphasizes the importance of a
topological approach of coarsening~\cite{Voorhees-EPL09}.

{\it Viscosity contrast induces topological symmetry breaking.} The
phase-separating melt under study is characterized by a strong
viscosity contrast. In the range of temperatures considered
[1160$^\circ$C-1210$^\circ$C], the viscosities are
$\eta_F \approx 10\; \mathrm{Pa.s}$ for the {\it fluid}
phase and $\eta_V \approx 10^6 \; \mathrm{Pa.s}$ for the {\it viscous}
phase~\cite{Bouttes-Thesis14,Bouttes-ActaMater15}.

As discussed above, coarsening relies on a succession of capillary
break-ups and retractions of ligaments. How is the dynamics affected
by the strong viscosity asymmetry?  From the early work of
Tomotika~\cite{Tomotika-PRSA35,*Tomotika-PRSA36} based on linear
stability analysis, to the more recent discussions about
end-pinching~\cite{Stone-JFM86,Stone-JFM89,Lister-PF98,Leal-PF01,Tong-PF07,Quan-JFM09}
and detailed studies of the
break-up~\cite{Cohen-PRL99,Zhang-PRL99,Lister-JFM03,Verdier-PRE07,Papageorgiou-IMAJAM13,Basaran-PNAS15},
the question of the stability of a viscous thread or droplet suspended
in another viscous fluid has raised a wide interest in fluid mechanics
(see \emph{e.g.} ref.~\cite{Eggers-RPP08} for a review). The present
situation is clearly more difficult to handle for at least two
reasons. First, the geometry of the inter-connected cluster and its ligaments is far more complex than a simple droplet or a
thread. In particular different curvatures have to be considered. A
closer configuration in this respect may be that of a toroidal droplet
~\cite{Pairam-PRL09,Mehrabian-JFM13}. Second, the liquid
surrounding the ligament about to break up is not homogeneous but is
itself a mixture of the two phases. An effective viscosity of the viscous
matrix can be obtained using a homogenization approach. However, long-range hydrodynamic interactions are likely to be
at play and introduce correlation between successive breaking events.

Forgetting about most of the complexity of the problem, we draw
simple scaling arguments, and estimate the time scales associated with
the two mechanisms at play: break-up and retraction of a ligament. The
only length scale to be considered here is that of the coarsening
$\ell(t)$. 

{\it Retraction} \--- A typical estimate for the retraction time $\tau_R$ of a ligament of size $\ell$ and viscosity $\eta_T$ suspended in a fluid of viscosity $\eta_M$ is~\cite{Sarkar-JFM01,Jackson-JRheol03}:
\begin{equation}
\tau_R \approx \frac{3\ell}{4}\frac{\eta_T+\eta_M}{\gamma}\;.
\label{retraction-time}
\end{equation}
For a disordered matrix surrounding the ligament, a simple estimate of the
effective viscosity gives $\eta_{M}\approx \phi\eta_F + (1-\phi)\eta_V
\approx (1-\phi)\eta_V$ \cite{Torquato-book02}. We thus obtain the
retraction times of a fluid and a viscous ligament respectively:
\begin{equation}
\tau_R^F \approx \frac{3}{4}(1-\phi)\frac{\eta_V\ell}{\gamma}\quad , \qquad
\tau_R^V \approx \frac{3}{4}(2-\phi)\frac{\eta_V\ell}{\gamma}\;.
\label{retraction-time-fluid-viscous}
\end{equation}
Despite the strong viscosity contrast the time scale is thus almost
the same for the retraction of a fluid or a viscous
ligament in a phase-separated melt.

{{\it Break-up} \--- Generalizing the time scale for the final break-up  
of a ligament~\cite{Eggers-PRL93,Brenner-PF96} to the case of viscosity contrast~\cite{Cohen-PRL99,Papageorgiou-IMAJAM13}, we get
\begin{equation}
\tau_R \approx H_0^{E}\frac{\eta_M\ell}{\gamma} \left( \frac{\eta_T}{\eta_M} \right)^{\alpha}\;,
\label{breaking-time}
\end{equation}
where again $\eta_M$ is the viscosity of the suspending fluid and
$\eta_T$ that of the ligament, the constant $H_0^E\approx
0.03$~\cite{Eggers-PRL93,Brenner-PF96} and the exponent $\alpha\approx
0.5-0.6$~\cite{Cohen-PRL99,Papageorgiou-IMAJAM13}. For simplicity, in
the following we consider $\alpha=0.5$ and get respectively for the
break-up time of a fluid and a viscous ligament surrounded
by the phase-separated mixture:
 \begin{equation}
\tau_B^F \approx \frac{\sqrt{1-\phi}}{H_0^E}\frac{\sqrt{\eta_V\eta_F}\ell}{\gamma}, \qquad
\tau_B^V \approx \frac{\sqrt{1-\phi}}{H_0^E}\frac{\eta_V\ell}{\gamma}\;.
\label{breaking-time-fluid-viscous}
\end{equation}
Unlike for retraction, the break-up time scale appears to be much
lower for a fluid ligament than for its viscous
counterpart. A first consequence is that ruptures of capillary bridges,
as shown in Fig.~\ref{pinch-off}, are much more likely in the
 fluid phase than in the viscous phase. Conversely, hole
resorptions are more likely in the {\it viscous} phase. A second
consequence is that the breaking time scale of a {\it fluid}
ligament may also become much lower than the retraction time. This
leaves room for an end-pinching like mechanism in the fluid phase,
\emph{i.e.} an additional rupture event during the retraction of a broken
capillary bridge, leading to fragmentation. In contrast, we expect
fragmentation to be very unlikely in the {\it viscous} phase.}

Following these simple scaling arguments, viscosity contrast should
induce a topological symmetry breaking during
viscous coarsening. This is indeed the case. In the supplementary
material (available on line) we show a movie corresponding to the
series of 3D tomography images of the coarsening
of glass G$_2$. Only the minority fluid phase experiences
fragmentation. 

\begin{figure}
	\centerline{\includegraphics[width=0.99\columnwidth]{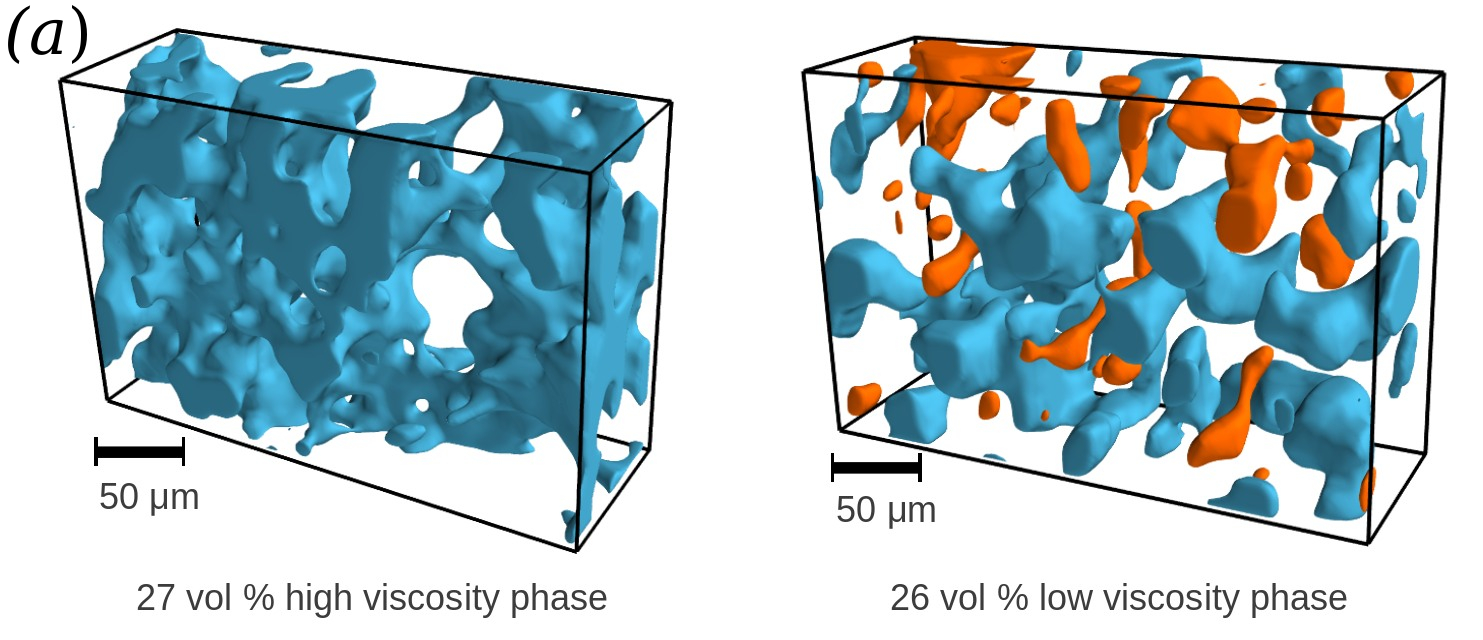}}
\centerline{\includegraphics[width=0.6\columnwidth]{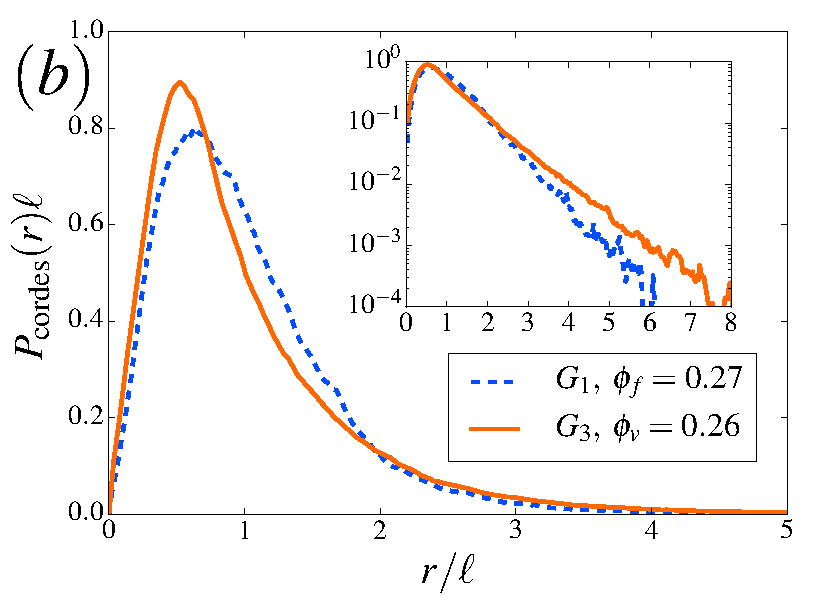}}
%
\caption{(a) Morphology of the minority phase after 20 min at 1150$^\circ$C
  for glasses G$_3$ (Left) and G$_1$ (Right). The largest connected
  domain is represented in blue; isolated domains in orange. When the
  minority phase is the {\it fluid} one (G$_1$, $\Phi=0.27$) a lot of
  isolated domains are present. This fragmentation process is absent
  when the minority phase is the {\it viscous} one (G$_3$,
  $\Phi=0.72$). (b) Chord distributions for the interconnected
  domains of (a), renormalized by the
  characteristic scale $\ell$. Semilog plot shown in inset.}
\label{Frag-vs-compo}
\end{figure}

Another spectacular illustration of this topological symmetry breaking is
given in Fig.~\ref{Frag-vs-compo}(a) that shows the contrasting
morphology of glasses G$_1$ and G$_3$ after 20 min of heat treatment. The two compositions ``mirror'' each other:
in G$_1$ the volume fraction of the {\it fluid} phase is $\Phi_1=0.27\pm
0.02$ while it is $\Phi_3= 0.72\pm 0.02\approx 1-\Phi_1$ in G$_3$. Here
only the minority phases are represented, respectively the {\it fluid}
phase for G$_1$ and the {\it viscous} phase for G$_3$. When the minority
phase is the fluid one, a significant fragmentation is observed. Many
isolated domains (here colored in orange) can be identified together with
the percolating domain (in blue). In the complementary case, no
fragmentation is observed in the minority {\it viscous}
phase~\cite{Eggers-JFM05}, only a fully connected domain is obtained. The
shape of the interconnected cluster also differs between G$_1$ and G$_3$: when
composed of the fluid phase, it has a more compact and globular
shape, while elongated structures are observed for the
viscous phase in G$_3$. In Fig.~\ref{Frag-vs-compo} (b), a
quantitative view on this morphological contrast is given by the
chord-length distribution~\cite{Torquato1993} (the histogram
of intercept lengths through the minority phase), with a more
heterogeneous distribution and a longer tail for the \emph{viscous}
phase (G$_3$).

{\it End pinching} \--- While the simple scenario of end-pinching
presented above is appealing, the actual fragmentation mechanisms at play
are likely to be more complex and interdependent due to hydrodynamic
interactions. Yet, another phenomenon may result
even more directly from end pinching: the refragmentation of already
fragmented isolated domains. Although the geometry of these domains is
not as well-controlled as that of the elongated drops used in the
experimental fluid mechanics set-ups, we benefit here from a population
of refragmenting drops significant enough for a statistical study.
According to the end-pinching scenario~\cite{Stone-JFM86,Stone-JFM89} an
elongated droplet that relaxes can either relax to a sphere, or fragment
in two or more droplets depending on the viscosity contrast and the
aspect ratio. 

\begin{figure}
\includegraphics[width=0.95\columnwidth]{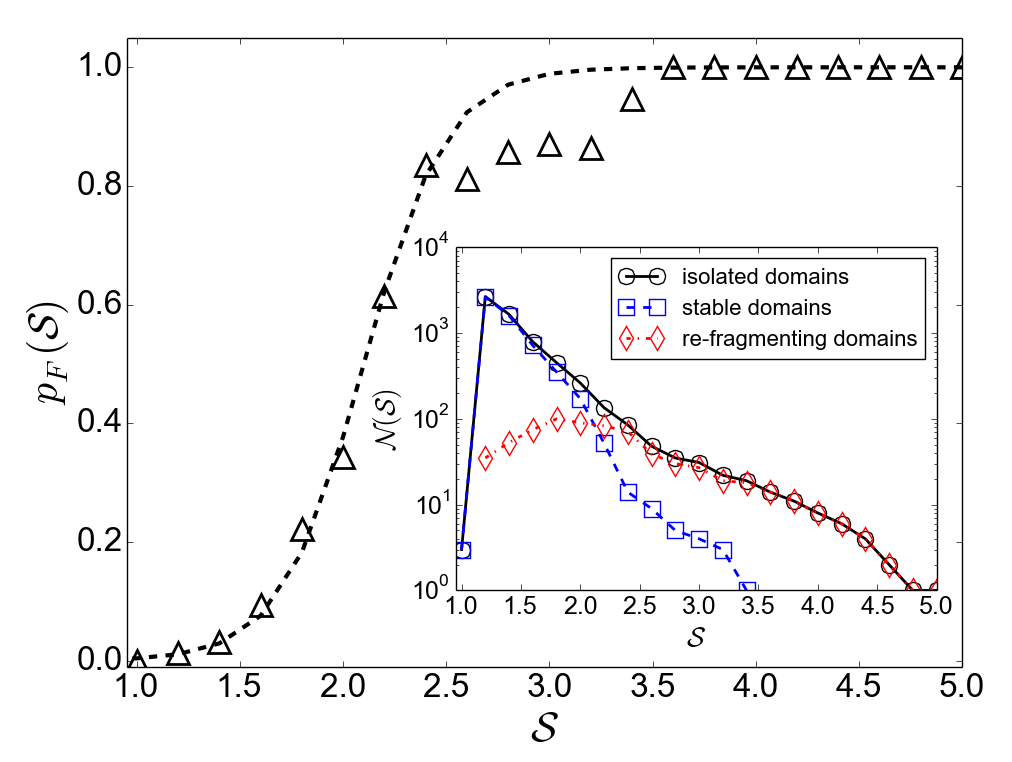}
\caption{Re-fragmenting probability of an isolated domain w.r.t. its
  sphericity index at first fragmentation (symbols). The dashed line
  shows a sigmoidal function $f(\mathscr{S})= 1/(1+e^{(\mathscr{S}-\mathscr{S}_c
  )/w_c})$ with $\mathscr{S}_c=2.1$ and $w_c=0.2$. Inset:
  Histograms of the sphericity index $\mathscr{S}$ at fragmentation of
  isolated domains (all, stable, refragmenting) of the three glasses
  G$_1$, G$_2$ and G$_3$. }
\label{Frag-vs-forme}
\end{figure}

The ratio of viscosity between a {\it fluid} droplet and the
phase-separated mixture is low enough here to
allow fragmentation. Still, a minimum extension is required for
rupture to take place: no breaking event takes place below a threshold
aspect ratio $L/R \lesssim 6$ where $L$ is the length of the elongated
drop before relaxation and $R$ is a typical
radius~\cite{Stone-JFM86,Stone-JFM89}. The end-pinching scenario can
be thus tested quantitatively via its dependence on the aspect ratio.
In Fig.~\ref{Frag-vs-forme} we reported all fragmentation events from
the interconnected cluster taking place along the coarsening of glasses
G$_1$, G$_2$ and G$_3$. In lieu of a ratio $L/R$ difficult to define
on our domains we use a simple sphericity index
$\mathscr{S}=S^{3/2}/6\sqrt{\pi}V$ where $S$ and $V$ are the surface and
volume of the isolated domain just after it has detached from the
infinite cluster. A spherical drop is such that $\mathscr{S}=1$ and an
elongated domain is such that $\mathscr{S}\gg 1$.  For a cylindrical
drop of aspect ratio $L/R=6$, corresponding to the stability threshold
of end-pinching, $\mathscr{S}_c \approx 4/3$. In the
inset of Fig.~\ref{Frag-vs-forme} we show the sphericity histogram for
the whole set of isolated domains at fragmentation as well as its
restrictions to domains that do or do not refragment
afterwards.  The two populations are clearly distinguishable. In the
main panel we show the cumulative probability of refragmenting
versus sphericity. We get a sharp transition at
$\mathscr{S}_c\approx 2.1\pm0.2$, a value a bit larger than for simple elongated droplets.

The scenario of end-pinching thus seems to reasonably account for the
refragmentation of isolated domains. Despite the complexity of the
material and the high-temperature conditions, these results support
the idea that the viscous coarsening of silicate melts is
fully accounted for by Newtonian fluid dynamics.

{\it Conclusion} \--- The dynamics of viscous coarsening appears to
be dramatically modified by a strong viscosity contrast between the
phases. A topological symmetry breaking is evidenced:
elementary breaking events occur only in the most fluid phase and lead
to fragmentation while resorption events occur only in the most
viscous phase that remains inter-connected. The crucial importance of
pinch-off mechanisms is illustrated by an end-pinching
fragmentation taking place in the fluid phase.

{\bf Acnowledgements} X-ray tomography experiments were performed on
beamline ID19 (proposals HD-501, SC-3724, LTP MA-1876). We thank Elodie
Boller, Davy Dalmas, Pierre Lhuissier and Luc Salvo for their
experimental help, Laurent Limat for interesting discussions, and Alban
Sauret for helpful comments and suggestions. Support of ANR project EDDAM
ANR-11-BS09-027 is acknowledged.


%

\end{document}